\documentclass[italian,10pt,reqno]{amsart}

\usepackage{amsmath,amssymb}
\usepackage{latexsym}
\usepackage{epsfig}
\usepackage{cite}
\usepackage{booktabs}
\usepackage{amsthm}
\usepackage{tikz,tikz-3dplot}
\usepackage{pgfplots}
\usepackage{pgf}

\usetikzlibrary{arrows,decorations.pathmorphing,backgrounds,positioning,fit,petri,decorations}
\usetikzlibrary{calc,intersections,through,backgrounds,mindmap,patterns,fadings}
\usetikzlibrary{decorations.text}
\usetikzlibrary{decorations.fractals}
\usetikzlibrary{fadings}
\usepackage{pgflibraryshapes}
\usetikzlibrary{lindenmayersystems}
\usetikzlibrary{shadings}
\usetikzlibrary{shadows}
\usetikzlibrary{shapes.geometric}
\usetikzlibrary{shapes.callouts}
\usetikzlibrary{shapes.misc}
\usetikzlibrary{spy}
\usetikzlibrary{topaths}
\usetikzlibrary{datavisualization}
\usetikzlibrary{datavisualization.formats.functions}

\newcommand{\bes}{\begin{equation*}}
\newcommand{\ees}{\end{equation*}}

\newtheorem{definition}{Remark}

\numberwithin{equation}{section}

\begin{document}

\title{Uniqueness of Segal quantization for oscillating systems}
\author{M. Bertini$^{1}$, S.L. Cacciatori$^{2,3}$ and M. Falchi Perna$^{2}$}
\address{\noindent $^1$Dipartimento di Matematica, Universit\`a degli Studi di Milano, Via Saldini 50, 20133 Milano, Italy\endgraf
$^2$Department of Science and High Technology, Universit\`a dell'Insubria, Via Valleggio 11, IT-22100 Como, Italy\endgraf
$^3$INFN sezione di Milano, via Celoria 16, IT-20133 Milano, Italy}

\normalsize
\maketitle
\pagestyle{myheadings}

\begin{abstract}
We show that the Segal quantization of an arbitrary system of decoupled harmonic oscillators is unique in the sense that the one particle Hilbert space is completely determined by the requests of being a naturally 
complex symplectic space carrying a unitary realization of the dynamical evolution of the considered system.
\end{abstract}

\section{Introduction}
The Segal quantization of an Hamiltonian system consists essentially in associating to the phase space a one particle states real Hilbert space $(\mathcal F, (\ ,\ ))$ bringing a symplectic structure $\omega^2$ and a complex structure $J$ such that
the complexification $H$ of $\mathcal F$ under $J$ has a complex scalar product
\begin{align}
 (\ ,\ )_{\mathbb C}=(\ ,\ )+i\omega^2
\end{align}
and the Hamiltonian evolution of the system is expressed by a unitary flow. Se for example \cite{Arnold,BaSeZh,Co-2,DeGe,Fr,Se-2,Sh}. The problem of the existence and uniqueness of the quantization for linear bosonic and fermionic fields 
has already been addressed in  \cite{Se-1,Se-3,Se-4,Se-5,We} and \cite{Ka}. Applications can be found in \cite{Co-1,De,Se-2,Sh,Weiss}. \\ 
The aim of the present paper is to provide a simple and direct proof of the existence and 
uniqueness of the Segal quantization 
for an arbitrary system, finite or infinite, of free harmonic oscillators, passing through a very explicit construction of the naturally complex symplectic space of one particle states, and the corresponding Hamiltonian operator realizing the 
unitary flux associated to the dynamical evolution.

\

\section{Naturally complex symplectic spaces}

For us, a symplectic space will be a triple $(\mathcal F,\omega^2,(\ ,\ ))$, with $\mathcal F$ a real Hilbert space endowed with a scalar product $(\ ,\ )$, and $\omega^2$ an antisymmetric non degenerate continuous bilinear form.
We will simply say that $\mathcal F$ is a symplectic space.

Let $I$ be the natural isomorphism between $\mathcal F$ and its dual $\mathcal F^*$ induced by $\omega^2$ along the definition
\begin{align}
\omega^2(x,Iy^*)=<y^*,x> \ \ \ 
\forall x\in\mathcal F,\ \forall y^*\in\mathcal F^*. 
\end{align}
In the finite dimensional case such isomorphism does not depend at all from the choice of a scalar product since all linear functionals over $\mathcal F$ are always continuous.

A vector field $V:D(V)\subset\mathcal F\rightarrow\mathcal F$ is called hamiltonian if there exists a function $\mathbf H:\mathcal F\rightarrow\mathbb{R}$ differentiable in $D(V)$, called hamiltonian, such that
\begin{align}
V(x)=Id\mathbf H(x)\ \ \ \forall x\in D(V). 
\end{align}
One defines the Poisson brackets between two differentiable functions $\mathbf K$ and $\mathbf H$ as the derivative of $\mathbf K$ along the direction of the hamiltonian field generated by $\mathbf H$:
\begin{align}
\{\mathbf K,\mathbf H\}=d\mathbf K(Id\mathbf H). 
\end{align}

If $D(d\mathbf K)\subset\mathcal F$ and $D(d\mathbf H)\subset\mathcal F$ are the domains of $d\mathbf K$ and $d\mathbf H$ respectively, the Poisson brackets $\{\mathbf K,\mathbf H\}$ is defined in
$D(\{\mathbf K,\mathbf H\})=D(d\mathbf K)\bigcap D(d\mathbf H)$.
According to the definition of the isomorphism $I$, we have obviously
\begin{align}
\{\mathbf K,\mathbf H\}=\omega^2(Id\mathbf H,Id\mathbf K). 
\end{align}
If $\mathcal F$ has finite dimensions, also the definition of Poisson brackets is independent from the choice of a scalar product. 

Let now $T$ be the natural isomorphism between $\mathcal F$ and its dual $\mathcal F^*$ defined by the relation 
\begin{align}
<x^*,y>=(Tx^*,y)\ \ \ \forall x^*\in\mathcal F^*,y\in\mathcal F. 
\end{align}
Through $T$ and $I$ let us construct the natural automorphism $J$ over $\mathcal F$ so defined:
\begin{align}
J=TI^{-1}. 
\end{align}
One easily verifies that
\begin{align}
\omega^2(x,y)=(x,Jy). 
\end{align}
The operator $J$ is antiselfadjoint, t.i. $J^\dagger=-J$. It depends not only on the symplectic form but also from the scalar product. Finally, said $\nabla\mathbf H:=Td\mathbf H$ the gradient of $\mathbf H$, we have
\begin{align}
Id\mathbf H=J^{-1}\nabla\mathbf H. 
\end{align}

We will say that the symplectic space $(\mathcal F,\omega^2,(\ ,\ ))$ is {\em naturally complex} if $J^2=-1$, and when this is the case we will call $J$ the complex unity of the space $\mathcal F$. Indeed, in this case $\mathcal F$ can be 
complexified in a natural way by defining the multiplication of its elements by complex numbers according to the definition:
\begin{align}
(\alpha+i\beta) x:=\alpha x+\beta Jx, \qquad\ \alpha,\beta\in \mathbb R.
\end{align}
Such complexification is dictated by the natural structure of $\mathcal F$.

In the finite dimensional case we call {\em $n-$dimensional standard symplectic space} the triple $(\mathcal F_S^n,\omega^2_S,(\ ,\ )_S)$ with $\mathcal F_S^n=\mathbb{R}^n_p\times\mathbb{R}^n_q$, $\omega^2_S=dp\wedge dq$ and
$(\ ,\ )_S$ the euclidean scalar product of $\mathbb{R}^{2n}$. This symplectic space is naturally complex, in this case being the imaginary unit $J_S$ represented by the matrix 
\begin{equation}
J_S=
\left(
\begin{array}{cc}
0 & 1\\
-1 & 0
\end{array}
\right).
\end{equation}
We will call $J_S$ the standard imaginary unit.

Two naturally complex symplectic spaces $(\mathcal F_1,\omega^2_1,(\ ,\ )_1)$ and $(\mathcal F_2,\omega^2_2,(\ ,\ )_2)$
will be said isomorphic if there exists a bijection $U$ between $\mathcal F_1$ and $\mathcal F_2$ that is simultaneously symplectic and isometric.
Let us give an example of a pair of isomorphic naturally complex symplectic spaces, which will be useful later on.
Set $(\mathcal F,\omega^2,(\ ,\ ))$ where $\mathcal F=\mathcal F_S^1$, $\omega^2(X,Y)=\omega^2_S(X,Y)$, $(X,Y)=(X,GY)_S$ with
\begin{equation}
G=
\left(
\begin{array}{cc}
\Omega^{-1} & 0\\
0 & \Omega
\end{array}
\right)
\end{equation}
and $\Omega>0$. It is easy to show that this space is symplectic and naturally complex and the complex unity is
\begin{equation}
J=
\left(
\begin{array}{cc}
0 & -\Omega\\
\Omega^{-1} & 0
\end{array}
\right).
\end{equation}
Such space is isomorphic to the standard symplectic space. Indeed, let $U(\alpha)$ be the one parameter group of biunivocal transformations between $\mathcal F_S$ and $\mathcal F$
\begin{align}
U(\alpha):\mathcal F_S\to\mathcal F 
\end{align}
defined by
\begin{equation}
U(\alpha)=
\left(
\begin{array}{cc}
\Omega^{1/2}\cos\alpha & -\Omega^{1/2}\sin \alpha\\
\Omega^{-1/2}\sin\alpha & \Omega^{-1/2}\cos\alpha
\end{array}
\right).
\end{equation}
It is clear that for each $\alpha$ it holds
\begin{equation}
\begin{array}{l}
(X,Y)=(U(\alpha)x,U(\alpha)y)=(x,y)_S\\
\omega^2(X,Y)=\omega^2(U(\alpha)x,U(\alpha)y)=\omega^2_S(x,y),
\end{array}
\end{equation}
so the two spaces are isomorphic.\\
We will now apply this formalism to the quantization of an arbitrary configuration of free harmonic oscillators, starting from the simplest case of a one dimensional harmonic oscillator.

\section{The one-dimensional harmonic oscillator}

On the standard naturally complex symplectic space $(\mathcal F_S^1,\omega^2_S,(\ ,\ )_S)$ let us consider the hamiltonian 
\begin{align*}
\mathbf K(X)=\frac{1}{2}\|\sqrt KX\|^2_S=\frac{1}{2}(P^2+\Omega^2Q^2) 
\end{align*}
where
\begin{equation}
K=
\left(
\begin{array}{cc}
1 & 0\\
0 & \Omega^2
\end{array}
\right),
\end{equation}
and $\Omega>0$.
The corresponding hamiltonian system is given by the equations
\begin{align*}
\dot X=I_Sd\mathbf K(X)=-J_S\nabla\mathbf K(X)=AX 
\end{align*}
with $A:\mathcal F_S\rightarrow\mathcal F_S$,
\begin{equation}
\label{system} 
A
\left(
\begin{array}{c}
P\\Q
\end{array}
\right)=
\left(
\begin{array}{cc}
0 & -\Omega^2 \\
1 & 0
\end{array}
\right)
\left(
\begin{array}{c}
P\\Q
\end{array}
\right).
\end{equation}
It is easy to see that if $\Omega\neq 1$ the hamiltonian flux $\phi_t=e^{At}$ generated by the previous equations is not unitary. We want to define a new naturally complex symplectic space $(\mathcal F,\omega^2,(\ ,\ ))$ 
with respect to which 
\begin{itemize}
\item [a)] 
the flow $\phi_t=e^{At}$ should be hamiltonian and unitary;
\item[b)] the variables $P$ and $Q$ should satisfy the canonical commutation relations.
\end{itemize}
\begin{definition}\label{def1}
 From now on we will say that a naturally complex symplectic space satisfying such conditions constitutes a unitary realization of the dynamics.
\end{definition}
Let us prove that in this case a unitary realization of the dynamics exists and is essentially unique.

\

Set $\mathcal F=\mathcal F_S$ and like before let $(\ ,\ )_S$ be the euclidean scalar product on $\mathbb{R}^2$. Then, 
$(\ ,\ )$, when it exists, can be written in the form $(X,Y)=(X,GY)_S$ with $G$ a symmetric matrix:
\begin{equation}
G=
\left(
\begin{array}{cc}
a & b\\
b & c
\end{array}
\right).
\end{equation}
For $\phi_t$ to be a group of unitary matrices with $t$ real, $A$ must be antisymmetric:
\begin{align*}
(X,AY)=-(AX,Y), \ \ \ \forall X,Y\in\mathcal F. 
\end{align*}
Thus, $G$ must have the form
\begin{equation}
G=
\left(
\begin{array}{cc}
\alpha & 0\\
0 & \Omega^2\alpha
\end{array}
\right)
\end{equation}
with $\alpha>0$.

Now, let $\omega^2$ be a symplectic form over $\mathcal F$. It can always be written in the form
\begin{align*}
\omega^2(X,Y)=(X,JY) 
\end{align*}
with $J$ antisymmetric. Set
\begin{equation}
J=
\left(
\begin{array}{cc}
a & b\\
c & d
\end{array}
\right),
\end{equation}
the condition
\begin{align*}
(X,JY)=-(JX,Y) \ \ \ \forall X,Y\in\mathcal F 
\end{align*}
implies 
\begin{equation}
J=
\left(
\begin{array}{cc}
0 & \Omega^2\beta\\
-\beta & 0
\end{array}
\right),
\end{equation}
with $\beta\neq 0$.
Since we must have $J^2=-1$, we get
\begin{equation}
J=J_\pm=\pm
\left(
\begin{array}{cc}
0 & \Omega\\
-\Omega^{-1} & 0
\end{array}
\right)
\end{equation}

Now, let us impose the canonical commutation relations to be satisfied. It is known that 
\begin{align}
\{Q,P\}=1,\ \ \ \{P,P\}=\{Q,Q\}=0,
\end{align}
corresponds to
\begin{align*}
\omega^2=dP\wedge dQ. 
\end{align*}
After imposing
\begin{align*}
dP\wedge dQ(X,Y)=\omega^2(X,Y)=(X,JY)\ \ \ \forall X,Y\in\mathcal F,
\end{align*}
we get $J=J_-$ and $\alpha=\Omega^{-1}$.
This way, also the scalar product is completely fixed:
\begin{align*}
((P,Q),(P',Q'))=\Omega^{-1}PP'+\Omega QQ'. 
\end{align*}

Let us verify that the field $V(X)=AX$ is hamiltonian. Since the field $AX$ is linear in $x$, if there exists an hamiltonian $\mathbf H:\mathcal F\rightarrow \mathbb{R}$ such that $AX=Id\mathbf H(X)$,
then it must hold true
\begin{align}
AX=-J\nabla\mathbf H(X)=-JHX,
\end{align}
which gives
\begin{equation}
H=JA=
\left(
\begin{array}{cc}
\Omega & 0\\
0 & \Omega
\end{array}
\right).
\end{equation}
Obviously $H=H^\dagger$, $[H,J]=0$. Thus, we can choose the hamiltonian 
\begin{align}
\mathbf H(X)=\frac{1}{2}\|\sqrt HX\|^2=\frac{1}{2}(P^2+\Omega^2Q^2). 
\end{align}
Notice that it coincides with the $\mathbf K(X)$ we used to define the hamiltonian field $AX$ in the standard symplectic space. 

Let us apply the canonical and symplectic transformation 
\begin{align}
U(\alpha):(\mathcal F_S,\omega^2_S,(\ ,\ )_S)\to(\mathcal F,\omega^2,(\ ,\ )) 
\end{align}
defined above, with $\alpha=0$:
\begin{align}
\begin{array}{l}
p=\Omega^{-1/2}P,\\
q=\Omega^{1/2}Q.
\end{array} 
\end{align}
The hamiltonian $\mathbf H(x)$ in the new variables is thus
\begin{align}
\mathbf H(x)=\frac{1}{2}\Omega(p^2+q^2),
\end{align}
and the equations of motion are
\begin{align}
\dot x=-J_S\nabla H(x)=-J_S\Omega x.
\end{align}
Thinking to $(\mathcal F_S,\omega^2_S,(\ ,\ )_S)$ as a naturally complex space, the considered equations of motion become
\begin{align}
\dot x=-i\Omega x, 
\end{align}
and the corresponding evolution group is
\begin{align}
\phi(t)=e^{-i\Omega t}. 
\end{align}

\section{System of a finite number of decoupled harmonic oscillators with different frequencies} 
In order to understand how Segal quantization works in general, the next step is to consider the case of a finite number of non interacting oscillators, avoiding degeneracies.
On the standard naturally complex symplectic space $(\mathcal F^n_S,\omega^2_S,(\ ,\ )_S)$ let us consider the hamiltonian system having hamiltonian
\begin{align*}
\mathbf K(x)=\frac{1}{2}\|\sqrt K\|^2_S=
\frac{1}{2}\sum_{i=1}^n(P_i^2+\Omega^2_iQ_i^2) 
\end{align*}
where
\begin{equation}
K=
\left(
\begin{array}{cc}
1 & 0 \\
0 & \Omega^2
\end{array}
\right),
\end{equation}
being $\Omega$ defined by its action on the standard basis $\{e_i\}_{i=1,\dots,n}$ of $\mathbb{R}^n$ as $\Omega e_i=\Omega_i e_i$, $\Omega_i>0,\  \Omega_i\neq\Omega_j\ \rm{per}\ i\neq j$. 
The corresponding hamiltonian equations are
\begin{align*}
\dot X=I_Sd\mathbf K(X)=-J_S\nabla\mathbf K(X)=AX 
\end{align*}
with $A:\mathcal F^n_S\rightarrow\mathcal F^n_S$,
\begin{equation}
\label{system_n} 
A
\left(
\begin{array}{c}
P\\Q
\end{array}
\right)=
\left(
\begin{array}{cc}
0 & -\Omega^2 \\
1 & 0
\end{array}
\right)
\left(
\begin{array}{c}
P\\Q
\end{array}
\right).
\end{equation}
Like for the one dimensional case, if $\Omega\neq 1$ then the hamiltonian flux $\phi_t=e^{At}$ generated by the previous equations is not unitary. Again, we will show that this dynamical system admits a unique unitary realization 
(see remark \ref{def1}).

\

Set $\mathcal F=\mathcal F_S$. After defining the scalar product over $\mathcal F$ by $(X,Y)=(X,GY)_S$ with $G^T=G$ let us impose for the operator $A$ to be antisymmetric:
$$
(X,AY)=-(AX,Y).
$$
Writing $G$ as 
\begin{equation}
\label{system_n} 
G=
\left(
\begin{array}{cc}
L & M \\
M^T & N
\end{array}
\right)
\end{equation}
with $L=L^T$, $N=N^T$, the antisymmetry of $A$ imposes
\begin{align}
M=-M^T,\ \ \Omega^2M=M\Omega^2,\ \ N=\Omega^2L.
\end{align}
Since $M$ is antisymmetric, then $iM$ is selfadjoint in $\mathbb C^n$ with the standard scalar product. Since $iM$ commutes with $\Omega^2$ they must have common eigenvectors. 
Since the eigenvectors of $\Omega^2$ are simple, then $iM$ is a real function of $\Omega$, $iM=f(\Omega)$. 
Hence, $M=-i(iM)=if(\Omega)$ is a matrix having only imaginary components and thus it cannot be a linear operator on $\mathbb R^n$ unless $f=0$. Thus we get $M=0$.
The fact that $N=\Omega^2L$ with $N^T=N$ implies that $L$ commutes with $\Omega^2$ and for the same reasons as before we deduce that $L=f(\Omega)$ with $f$ a real and positive function (since the scalar product $(\ ,\ )$ 
must be positive). In conclusion, the most general scalar product over $\mathcal F$ with respect to which $A$ is antisymmetric is:
\begin{align*}
((P,Q),(P',Q'))=(P,f(\Omega)P')_S+(Q,\Omega^2f(\Omega)Q')_S 
\end{align*}
with $f$ a positive real function.

A symplectic form $\omega^2$ over $\mathcal F$ can always be written in the form
\begin{align}
\omega^2(X,Y)=(X,JY) 
\end{align}
with $J$ antisymmetric. Writing $J$ as
\begin{equation}
J=
\left(
\begin{array}{cc}
O & T \\
S & R
\end{array}
\right),
\end{equation}
the condition
\begin{align*}
(X,JY)=-(JX,Y)\ \ \ \forall X,Y\in\mathcal F 
\end{align*}
implies 
\begin{equation}
\begin{array}{l}
f(\Omega)O=-O^Tf(\Omega)\\
\Omega^2f(\Omega)R=-R^T\Omega^2f(\Omega)\\
f(\Omega)T=-S^T\Omega^2f(\Omega).
\end{array}
\end{equation}
If we interpret the operator $O$ as an operator over the vector space $\mathbb{R}^n_P,\ (\cdot,f(\Omega)\cdot)$, the operator $R$ as an operator over the vector space $\mathbb{R}^n_Q,\ (\cdot,\Omega^2f(\Omega)\cdot)$, 
the operator $S$ as an operator from $\mathbb{R}^n_P,\ (\cdot,f(\Omega)\cdot)$ to $\mathbb{R}^n_Q,\ (\cdot,\Omega^2f(\Omega)\cdot)$ and the operator $T$ as an operator from $\mathbb{R}^n_Q,\ (\cdot,\Omega^2f(\Omega)\cdot)$ to
$\mathbb{R}^n_P,\ (\cdot,f(\Omega)\cdot)$, then the three conditions above can be rewritten as
\begin{equation}
\begin{array}{l}
O^\dagger=-O\\
R^\dagger=-R\\
T^\dagger=-S.
\end{array}
\end{equation}
Let us now impose the conditions $\{Q_i,Q_j\}=\{P_i,P_j\}=0$. Said $e_1,\dots,e_n$ the standard basis of $\mathbb{R}^n$, such conditions are equivalent to
\begin{equation}
\begin{array}{l}
\omega^2((e_i,0),(e_j,0))=0\ \ \ \forall i,j,\\
\omega^2((0,e_i),(0,e_j))=0\ \ \ \forall i,j,
\end{array}
\end{equation}
which are satisfied if and only if $O=0$ e $R=0$. Therefore, the operator $J$  has the form 
\begin{equation}
J=
\left(
\begin{array}{cc}
0 & T \\
-T^\dagger & 0
\end{array}
\right).
\end{equation}
Now, let us impose the condition $\{Q_i,P_j\}=\delta_{i,j}$, which are equivalent to  
\begin{align}
\omega^2((e_i,0),(0,e_j))=\delta_{i,j}\ \ \ \forall i,j. 
\end{align}
By employing the properties of $J$ we then get 
\begin{align}
f(\Omega)T=1 
\end{align}
hence
\begin{align}
T=f(\Omega)^{-1}.
\end{align}
After imposing the natural complexity condition $J^2=-1$, we get $TT^\dagger=T^\dagger T=-1$, which leads to
\begin{align}
f(\Omega)=\Omega^{-1}. 
\end{align}
Thus, the symplectic form $\omega^2$ and the scalar product $(\ ,\ )$ have been completely fixed. In particular,
\begin{equation}
J=
\left(
\begin{array}{cc}
0 & \Omega\\
-\Omega^{-1} & 0
\end{array}
\right).
\end{equation}
In the same way as for the single one dimensional harmonic oscillator, we can easily show that the field $AX$ is hamiltonian, with
\begin{align*}
\mathbf H(X)=\frac{1}{2}\|\sqrt HX\|^2=\frac 12(P^2+\Omega^2 Q^2).
\end{align*}

Such hamiltonian coincides with the $\mathbf K(X)$ classically chosen in order to define the equations of motion of a system of oscillators on the standard naturally complex symplectic space.

Notice that the operator $J$ commutes with $H$.

Thus, we have proved that there exists one and only one naturally complex symplectic space with respect to which the evolution $\phi_t$ of a chain of decoupled harmonic oscillators with frequencies pair to pair distinct 
is unitary and hamiltonian. In this space the scalar product is given by 
\begin{align*}
((P,P'),(Q,Q'))=(P,\Omega^{-1}P')_S+(Q,\Omega Q')_S, 
\end{align*}
the symplectic form is the standard one
\begin{align*}
\omega^2=dP\wedge dQ 
\end{align*}
while the complex unit $J$ is given by
\begin{align*}
J(P,Q)=(\Omega Q,-\Omega^{-1}P),
\end{align*}
and the hamiltonian of the system is
\begin{align*}
\mathbf K(P,Q)=\frac{1}{2}\sum_{1=1}^n(P_i^2+\Omega_i^2Q_i^2). 
\end{align*}

We can finally consider the canonical and unitary transformation between the naturally complex spaces $\mathcal F,\omega^2,(\ ,\ )$ and $\mathcal F_S,\omega^2_S,(\ ,\ )_S$:
\begin{align}
\begin{array}{l}
p = \Omega^{-1/2}P\\
q = \Omega^{1/2}Q,
\end{array} 
\end{align}
So that the hamiltonian in the new variable is
\begin{align}
\mathbf H(x)=\frac 12(\Omega p,p)_S+(\Omega q,q)_S). 
\end{align}
and the equations of motion are
\begin{align}
\dot x=-J_S
\left(
\begin{array}{cc}
\Omega & 0\\
0 & \Omega
\end{array}
\right)x=-J_S\overline\Omega x 
\end{align}
Looking at the space $(\mathcal F_S,\omega^2_S,(\ , \ )_S)$ as a complex space, the last equations are equivalent to
\begin{align}
\dot x=-i\overline\Omega x,
\end{align}
and the flow they generate is
\begin{align}
\phi(t)=e^{-i\overline\Omega t}. 
\end{align}

\section{System of an infinite number of decoupled harmonic oscillators with different frequencies}
We can further improve our construction by passing to the case of infinite harmonic oscillators, yet avoiding degeneracies. 
Let us consider the measurable space $(X,\mu)$, $X\subset[0,+\infty)$, $\mu=\mu_{ac}+\mu_d$, $\mu_{ac}$ absolutely continuous, $\mu_d$ discrete. Let $\mathcal M$ be the space of all real functions over $X$,
$f=f(k)$, that are measurable with respect to $\mu$. In particular, let us denote with $s$ the function $s(k)=1\ \forall k\in X$.

We denote with $\mathcal M_{s}$ the standard Hilbert space $L^2(X,d\mu)$ and with $(\ ,\ )_s$ its standard scalar product. Given a function $\theta\in\mathcal M$, $\theta>0\ q.o.$ we call $\mathcal M_{\theta}$ the Hilbert space
\begin{align*}
\mathcal M_{\theta}=\{\forall f\in\mathcal M:\ \theta f\in\mathcal M_s\}, 
\end{align*}
with scalar product $(\ ,\ )_\theta$ 
\begin{align*}
(f,g)_\theta=(\theta f,\theta g)_{s}.
\end{align*}

Given two functions $\rho,\sigma\in\mathcal M$ positive a.e., let $\mathcal F_{\rho,\sigma}=\mathcal M_\rho\times\mathcal M_\sigma$ be the Hilbert space of pairs $(p,q)$, $p\in\mathcal M_\rho$, $q\in\mathcal M_\sigma$ 
with scalar product $(x,y)_{\rho,\sigma}=(p,p)_\rho+(q,q)_\sigma$. Given over $\mathcal F_{\rho,\sigma}$ a symplectic form $\omega^2$, we will look at $\mathcal F_{\rho,\sigma}$ as a symplectic space.

Over the space $\mathcal F=\mathcal M\times\mathcal M$ let us consider the system of differential equations
\begin{equation}
\dot x=Ax,
\end{equation}
with $A:\mathcal F\rightarrow\mathcal F$,
\begin{equation}
\label{system} 
A
\left(
\begin{array}{c}
p\\q
\end{array}
\right)=
\left(
\begin{array}{cc}
0 & -\Omega^2 \\
1 & 0
\end{array}
\right)
\left(
\begin{array}{c}
p\\q
\end{array}
\right)
\end{equation}
where $\Omega:\mathcal M\rightarrow\mathcal M$ is the multiplication operator $(\Omega f)(k)=kf(k)$. 
We will show that this dynamical system admits a unique unitary realization $(\mathcal F,\omega^2,(\ ,\ ))$, which is topologically equivalent to some $\mathcal F_{\rho,\sigma}$.

First, notice that $\phi_t:\mathcal F\to\mathcal F$ is given by
\begin{equation}
\phi_t
\left(
\begin{array}{c}
p \\ q
\end{array}
\right)
=
\left(
\begin{array}{cc}
\cos kt & -k\sin kt \\ 
\displaystyle \frac{\sin kt}{kt} & \cos kt
\end{array}
\right)
\left(
\begin{array}{c}
p \\ q
\end{array}
\right).
\end{equation}
Let us look for which $\rho$ and $\sigma$ we have
\begin{align}
\phi_t:\mathcal F_{\rho,\sigma}\rightarrow\mathcal F_{\rho,\sigma}. 
\end{align}
Clearly $\phi_t(p_0,q_0)\in\mathcal F_{\rho,\sigma}$ for all $(p_0,q_0)\in\mathcal F_{\rho,\sigma}$ and for all $t\in\mathbb{R}$ if and only if
\begin{equation}
\begin{array}{ll}
k\sin (kt)\, q_0\in\mathcal M_{\rho} & \forall q_0\in\mathcal M_\sigma\\
\displaystyle\frac{\sin kt}{k}p_0\in\mathcal M_{\sigma} & 
\forall p_0\in\mathcal M_\rho
\end{array}
\end{equation}
that is
\begin{equation}
\label{phi_su_F}
\begin{array}{l}
k\sin (kt)\, \rho\sigma^{-1}\in L^\infty(X,d\mu),\\
\displaystyle\frac{\sin kt}{k}\sigma\rho^{-1}\in L^\infty(X,d\mu).
\end{array}
\end{equation}

Let us introduce over $\mathcal F_{\rho,\sigma}$ a scalar product $(\ ,\ )$ equivalent to $(\ ,\ )_{\rho,\sigma}$. Then, there must exist four continuous linear operators 
\begin{equation}
\begin{array}{l}
M:\mathcal M_\rho\rightarrow\mathcal M_\rho \\
N:\mathcal M_\sigma\rightarrow\mathcal M_\sigma\\
O:\mathcal M_\sigma\rightarrow\mathcal M_\rho\\
P:\mathcal M_\rho\rightarrow\mathcal M_\sigma\\
\end{array}
\end{equation}
satisfying the conditions
\begin{equation}
\begin{array}{l}
M^\dagger=M\ \ \ M>0\\
N^\dagger=N\ \ \ N>0\\
O^\dagger=P\\
P^\dagger=O,
\end{array}
\end{equation}
and such that
\begin{align}
(x,y)=(x,Gy)_{\rho,\sigma}
\ \ \ \forall x,y\in\mathcal F_{\rho,\sigma} 
\end{align}
with
\begin{equation}
G
\left(
\begin{array}{c}
p \\ q
\end{array}
\right)
=
\left(
\begin{array}{cc}
M & O\\
P & N
\end{array}
\right)
\left(
\begin{array}{c}
p \\ q
\end{array}
\right).
\end{equation}

We impose for $\phi_t$ to be unitary with respect to the scalar product$(\ ,\ )$. Therefore, the operator $A$ must be at least antisymmetric. Making explicit the condition
\begin{align*}
 (x,Ax')=-(Ax,x')
\end{align*}
where $x=(p,q),\ x'=(p',q')$, we get
\begin{equation}
\label{condizioni_su_G}
\begin{array}{l}
(p,Op')_\rho=-(p,Pp')_\sigma\\
(q,P\Omega^2 q')_\sigma=(\Omega^2 q,Oq')_\rho\\
(p,M\Omega^2 q')_\rho=(p,Nq')_\sigma\\
(q,Np')_\sigma=(\Omega^2 q,Mp')_\rho.
\end{array}
\end{equation} 
Looking at $O$ as an operator defined on $D(O)=\mathcal M_\sigma\cap\mathcal M_\rho\subset\mathcal M_\rho$ with values in $\mathcal M_\rho$ and denoting with $O^T$ its adjoint, the first of conditions \eqref{condizioni_su_G} becomes
\begin{align}
O^T=-O. 
\end{align}
Keeping into account the symmetry of the operator $\Omega$ independently from the space $\mathcal M_\rho$ or $\mathcal M_\sigma$, on which it is defined by the chain of identities 
\begin{align}
(q,P\Omega^2 q')_\sigma=(\Omega^2 Oq,q')_\rho=(q,O^T\Omega^2q')_\rho=
-(q,O\Omega^2q')_\rho,
\end{align}
the second of \eqref{condizioni_su_G} becomes
\begin{align}
[O,\Omega^2]=0. 
\end{align}
Therefore, $O$ is selfadjoint and it commutes with $\Omega^2$. If we look at $O$ as an operator over the space $\mathcal M_\rho^C$ of functions defined in  $X$ with values in $\mathbb C$ with scalar product $(f,g)^C_\rho=(f^*,g)_\rho$, 
$iO$ is selfadjoint and commute with $\Omega^2$. Since the proper and improper eigenspaces of $\Omega$ are simple, this ensures that the operator $iO$ is a real function of $\Omega$, that is $iO=\xi(k)$ with 
$\xi$ real. The operator $O=-i(iO)=-i\xi(k)$ is an operator defined on the real space $\mathcal M_\rho$ if and only if $\xi=0$. Therefore $O=0$.\\
Let us now consider the last two conditions in \eqref{condizioni_su_G}. We have
\begin{align*}
(p,M\Omega^2q')_\rho=(p,Nq')_\sigma=(\Omega^2p,Mq')_\rho=(p,\Omega^2Mq')_\rho, 
\end{align*}
from which
\begin{align*}
[M,\Omega^2]=0. 
\end{align*}
Since $M$ is selfadjoint and commutes with $\Omega^2$, for the same reasons as above we have that $M$ is a (real) function of $\Omega$, $M=\eta^2(k)$, $\eta>0\ q.o.$.
Exploiting the third condition
\begin{align*}
\int_X\rho(k)p(k)\rho(k)\eta(k)k^2q'(k)=
\int_X\sigma(k)p(k)\sigma(k)(Nq')(k) 
\end{align*}
we get that also $N$ is a multiplication operator 
\begin{align*}
(Nq)(k)=\rho^2(k)\sigma^{-2}(k)\eta^2(k)k^2. 
\end{align*}
Thus, we conclude that for the operator $A$ to be antisymmetric the space $\mathcal F=\mathcal F_{\rho,\sigma}$ must be of the form $\mathcal F_{\xi,\xi\Omega}$. 
Notice that on such spaces the conditions \eqref{phi_su_F}, necessary and sufficient for the flow $\phi_t$ to be defined on the whole space, are automatically satisfied. Therefore,
{\it a necessary condition for the flow $\phi_t$ to be defined on the whole space $\mathcal F=\mathcal F_{\rho,\sigma}$ and to be unitary is that $\sigma=\rho\Omega$}.

Let now $\omega^2$ be a symplectic form over $\mathcal F_{\rho,\rho\Omega}$. It can always be written in the form
\begin{align*}
\omega^2(x,y)=(x,Jy) 
\end{align*}
with $J$ antiselfadjoint. After writing $J$ as
\begin{equation}
J=
\left(
\begin{array}{cc}
O & Q \\
P & R
\end{array}
\right)
\end{equation}
with
\begin{equation}
\begin{array}{l}
O:\mathcal M_\rho\rightarrow\mathcal M_\rho\\
R:\mathcal M_{\rho\Omega}\rightarrow\mathcal M_{\rho\Omega}\\
Q:\mathcal M_{\rho\Omega}\rightarrow\mathcal M_\rho\\
P:\mathcal M_\rho\rightarrow\mathcal M_{\rho\Omega}
\end{array}
\end{equation}
continuous, the condition
\begin{align*}
(x,Jy)=-(Jx,y)\ \ \ \forall x,y\in\mathcal F 
\end{align*}
implies
\begin{equation}
\begin{array}{l}
O^\dagger=-O\\
R^\dagger=-R\\
Q^\dagger=-P.
\end{array}
\end{equation}
Let us impose for the Poisson brackets among the dynamical variables $p$ and $q$ to be canonical. This implies that the symplectic form $\omega^2$ coincides with the standard one. In particular, it must hold
\begin{equation}
\begin{array}{l}
\omega^2((p,0),(p',0))=0\ \ \ \forall p,p'\in\mathcal M_\rho\\
\omega^2((0,q),(0,q'))=0\ \ \ \forall q,q'\in\mathcal M_{\rho\Omega}.
\end{array}
\end{equation}
Using the properties of $J$, we see that the previous equations are satisfied if and only if $O=0$ e $R=0$. Therefore, the operator $J$ has necessarily the form
\begin{equation}
J=
\left(
\begin{array}{cc}
0 & Q \\
-Q^\dagger & 0
\end{array}
\right).
\end{equation}
Let us now impose the condition
\begin{align}
\omega^2((p,0),(0,q))=(p,q)_s. 
\end{align}
Using $J$ we get
\begin{align}
\rho^2Q=1 
\end{align}
from which
\begin{align}
Q=\rho^{-2}. 
\end{align}
We now impose $J^2=-1$, so that the space $\mathcal F_{\rho,\rho\Omega}$ is naturally complex. This implies $QQ^\dagger=Q^\dagger Q=-1$, which leads to
\begin{align}
\rho=\Omega^{-1/2}. 
\end{align}
Hence, the symplectic form $\omega^2$ and the scalar product $(\ ,\ )$ have been completely fixed.

Again, we get that the field $Ax$ is hamiltonian, with
\begin{align}
\mathbf H(x)=\frac{1}{2}\|\sqrt Hx\|^2,
\end{align}
where
\begin{equation}
H=
\left(
\begin{array}{cc}
\Omega & 0\\
0 & \Omega
\end{array}
\right).
\end{equation}
This hamiltonian coincides with the $\mathbf K(x)$ classically chosen to define the equations of motion of a system of oscillators on the standard naturally complex symplectic space.
It is worth to notice that the operator $J$ commutes with $H$
\begin{align*}
[J,H]=0. 
\end{align*}

Thus, we have proved that there exists one and only one naturally complex symplectic space with respect to which the evolution $\phi_t$ of a chain of decoupled harmonic oscillators with pair to pair different frequencies is unitary and hamiltonian.
In this space the scalar product is given by
\begin{align*}
((p,p'),(q,q'))=(p,\Omega^{-1}p')_s+(q,\Omega q')_s, 
\end{align*}
the symplectic form is the standard one
\begin{align*}
\omega^2((p,q),(p',q'))=(p,q')_s-(q,p')_s 
\end{align*}
while the complex unity $J$ is given by
\begin{align*}
J(p,q)=(\Omega q,-\Omega^{-1}p). 
\end{align*}
The hamiltonian of the system is 
$$
\mathbf H(p,q)=\frac{1}{2}(\|p\|^2_s+\|\Omega q\|^2_s).
$$

\section{System of decoupled harmonic oscillators with equal frequency}
We can now introduce degeneracies. 
On the standard naturally complex symplectic space $\mathcal F^1_S$ let us consider the hamiltonian
\begin{align*}
\mathbf K(x)=\frac 12\|\sqrt Kx\|^2_S=
\frac 12(p_1^2+p_2^2+\Omega^2q_1^2+\Omega^2 q_2^2) 
\end{align*}
where
\begin{equation}
K=
\left(
\begin{array}{cccc}
1 & 0 & 0 & 0\\
0 & 1 & 0 & 0\\
0 & 0 & \Omega^2 & 0\\
0 & 0 & 0 & \Omega^2
\end{array}
\right)
\end{equation}
and $\Omega>0$. The corresponding hamiltonian system is given by the equations
\begin{align*}
\dot x=I_Sd\mathbf K(x)=-J_S\nabla\mathbf K(x)=Ax 
\end{align*}
with $A:\mathcal F\to\mathcal F$,
\begin{equation}
A
\left(
\begin{array}{c}
p\\
q
\end{array}
\right)
=
\left(
\begin{array}{cc}
0 & -\Omega^2\\
1 & 0
\end{array}
\right)
\left(
\begin{array}{c}
p\\
q
\end{array}
\right).
\end{equation}
If $\Omega\neq1$ the flow $\phi^{At}$ generated by the previous equations is not unitary. Once again, we will show that this system admits a unique unitary realization $(\mathcal F,\omega^2,(\ ,\ ))$.

\

As before, let $(\ ,\ )_S$ be the euclidean scalar product in $L^2(X,d\mu)\times L^2(X,d\mu)$. Therefore, the scalar product $(\ ,\ )$, when it exists, can be written in the form $(\cdot , \cdot)=(\cdot , G\cdot)$, with
\begin{equation}
G=
\left(
\begin{array}{cc}
a & b\\
b & c
\end{array}
\right),
\end{equation}
where $a,b,c$ are positive defined symmetric operators. For $\phi_t$ to be a group of unitary transformations, $A$ must be antisymmetric:
\begin{align}
(x,Ay)=-(Ax,y)\ \ \forall x,y\in\mathcal F,
\end{align}
so that $G$ must have the form
\begin{equation}
G=
\left(
\begin{array}{cc}
\alpha & 0\\
0 & \Omega^2\alpha
\end{array}
\right)
\end{equation}
with $\alpha$ a symmetric and positive definite operator.

Any symplectic form $\omega^2$ over $\mathcal F$ can be written in the form 
\begin{align}
\omega^2(x,y)=(x,Jy) 
\end{align}
with $J$ antisymmetric. After setting
\begin{equation}
J=
\left(
\begin{array}{cc}
a & b\\
c & d
\end{array}
\right),
\end{equation}
the condition
\begin{align*}
(x,Jy)=-(Jx,y)\  \  \forall x,y\in\mathcal F 
\end{align*}
implies 
\begin{equation}
\label{con}
\begin{array}{l}
\alpha a=-(\alpha a)^T,\\
\alpha d=-(\alpha d)^T,\\ 
\Omega^2\alpha c=b^T\alpha
\end{array}
\end{equation}
that is
\begin{equation}
\label{con_i}
\begin{array}{l}
\alpha a\alpha^{-1}=-a^T,\\
\alpha d\alpha^{-1}=-d^T,\\ 
\Omega^2\alpha c\alpha^{-1}=b^T.
\end{array}
\end{equation}
From the first two we deduce that $a$ is similar to $-a^T$ and $d$ is similar to $-d^T$, which is possible if and only is both $a$ and $d$ are vanishing 
matrices.\footnote{The matrices $a$ and $-a^T$ are similar, hence they have the same eigenvalues; on the other side also the matrices $a$ and $a^T$ have the same eigenvalues.
Therefore, $a^T$ and $-a^T$ have the same eigenvalues, which is possible if and only if all eigenvalues vanish and the whole matrix $a$ vanishes. }
Therefore, 
\begin{equation}
J=
\left(
\begin{array}{cc}
0 & \Omega^2{\alpha^{-1}}^Tc^T\alpha^T\\
c & 0
\end{array}
\right).
\end{equation}
For $J$ to be non degenerate it occurs for $c$ to be invertible.

The condition $J^2=-1$, so that $J$ is an imaginary unit, leads to $bc=-1$  that is $c^Tb^T=-1$, or also
\begin{align}
c^T\alpha c\alpha^{-1}=-\Omega^{-2} 
\end{align}
and finally
\begin{align}
{\alpha^{-1}}^Tc^T\alpha^Tc=-\Omega^{-2} 
\end{align}
from which the matrix $J$ takes the form
\begin{equation}
J=
\left(
\begin{array}{cc}
0 & -c^{-1}\\
c & 0
\end{array}
\right).
\end{equation}

Let us impose the conditions
\begin{align}
\{p_i,q_j\}=\delta_{i,j} 
\end{align}
that is
\begin{equation}
\begin{array}{l}
\omega^2((e_i,0),(0,e_j))=\delta_{i,j}\\
\omega^2((0,e_i),(e_j,0))=\delta_{i,j}.
\end{array}
\end{equation}
These imply $\alpha=c$ and $\alpha=\pm\Omega^{-1}$. The positivity of the scalar product imposes $\alpha=\Omega^{-1}$.
Hence, the scalar product is finally
\begin{equation}
G=
\left(
\begin{array}{cc}
\Omega^{-1} & 0\\
0 & \Omega
\end{array}
\right),
\end{equation}
whereas the matrix $J$ is given by
\begin{equation}
J=
\left(
\begin{array}{cc}
0 & \Omega\\
-\Omega^{-1} & 0
\end{array}
\right).
\end{equation}
It is worth noticing that the operator $J$ commutes with $H$.

Thus, we have shown that there exists one and only one naturally complex symplectic space with respect to which the evolution $\phi_t$ of a chain of decoupled harmonic oscillators with all identical frequencies is unitary
and hamiltonian. In this space the scalar product is given by
\begin{align*}
((p,q),(p',q'))=(p,\Omega^{-1}p')_S+(q,\Omega q')_S, 
\end{align*}
the symplectic form is the standard one
\begin{align*}
\omega^2=dp\wedge dq 
\end{align*}
whereas the complex unity $J$ is given by
\begin{align*}
J(p,q)=(\Omega q,-\Omega^{-1}p). 
\end{align*}
The hamiltonian of the system is
\begin{align*}
\mathbf K(p,q)=\frac{1}{2}\sum_{1=1}^n(p_i^2+\omega_i^2q_i^2). 
\end{align*}
With respect to this scalar product and symplectic form, the hamiltonian of the system is given by
\begin{align*}
\mathbf H(p,q)=\frac{1}{2}(\|p\|^2_S+\|\Omega q\|^2_S).
\end{align*}
The flow of such hamiltonian system is by construction symplectic and unitary.

\

\section{The general case of a system of finite or infinite harmonic oscillators with arbitrary frequencies}
We can finally tackle the most general case.
Let us consider the measurable space $(X,\mu)$, $X\subset [0,\infty)\times \mathbb R$, $\mu=\mu_{ac}+\mu_d$. Let $\mathcal M$ be the space of real functions of real variables $f=f_\xi(k)$ that are measurable w.r.t. $\mu$. \\
Over the space $\mathcal F=\mathcal M\times \mathcal M$ let us consider the system of differential equations
\begin{align}
 \dot x=Ax
\end{align}
with $A: \mathcal F\rightarrow \mathcal F$,
\begin{align}
 A
\left(
\begin{array}{c}
p\\
q
\end{array}
\right)
=
\left(
\begin{array}{cc}
0 & -\Omega^2\\
1 & 0
\end{array}
\right)
\left(
\begin{array}{c}
p\\
q
\end{array}
\right),
\end{align}
where $\Omega: \mathcal M\rightarrow \mathcal M$ is the multiplication operator $(\Omega f)_\xi(k)=kf_\xi(k)$.  Using the previous results it is possible to prove that there exists exactly one unitary realization of this dynamical system.
Since it is a summary of the results in the previous sections, we leave the proof of this statement to the interested reader. This is the most general case for a positive adjoint operator in a Hilbert space, with arbitrary spectrum and arbitrary
degeneration.

\section{Segal quantization}
With all this at hand the Segal quantization of a system of harmonic oscillators is direct. Given the system of harmonic oscillators we construct the (unique) correspondent naturally complex symplectic space $\mathcal F$, endowed with the 
scalar product
\begin{align}
 ((p,q),(p',q'))=(p,\Omega^{-1}p')_S+(q,\Omega q')_S, 
\end{align}
the standard symplectic form 
\begin{align}
 \omega^2((p,q),(p',q'))=(p,q')_S-(q,p')_S,
\end{align}
and with the complex unit $J$ 
\begin{align}
 J(p,q)=(\Omega q, -\Omega^{-1}p).
\end{align}
The hamiltonian of the system is
\begin{align}
\mathbf H(p,q)=\frac{1}{2}(\|p\|^2_S+\|\Omega q\|^2_S),
\end{align}
and the dynamics is given by the unitary and symplectic one parameter group 
\begin{align}
U_t=e^{-iH t},
\end{align}
where $H$ is the linear self adjoint operator $H(p,q)=(\Omega p, \Omega q)$.\\
In order to quantize this system we consider on the symmetric Fock space 
\begin{align}
\Gamma(\mathcal F)=\mathbb C\oplus \mathcal F\oplus (\mathcal F\otimes_s \mathcal F) \oplus (\mathcal F\otimes_s \mathcal F\otimes_s \mathcal F)\oplus \ldots,
\end{align}
the unitary group of evolution
\begin{align}
 \mathbb U_t =\Gamma (U_t)
\end{align}
generated by the hamiltonian
\begin{align}
 \mathbb H=d\Gamma (H).
\end{align}
The remaining construction is the standard one with the formalism of annihilation and creation operators.

\

\section*{acknowledgment}
We thank A. Posilicano and D. Noja for helpful conversation.

\end{document}